# Hafnia based Phase-Change Ferroelectric Steep-Switching FETs on a 2-D MoS$_2$ platform

*Sooraj Sanjay$^‡$, Jalaja M. A$^‡$, Navakanta Bhat\*, and Pavan Nukala\**

Centre for Nano Science and Engineering, Indian Institute of Science, Bengaluru, India.
Corresponding author email: navakant@iisc.ac.in, pnukala@iisc.ac.in

## Abstract

Ferroelectric field-effect transistors integrated on 2D semiconducting platforms are extremely relevant for low power electronics. Here, we propose and demonstrate a novel phase-change ferroelectric field effect transistor (PCFE-FET) for steep switching applications. Our gate stack is engineered as a ferroelectric Lanthanum doped hafnium oxide (LHO) proximity coupled with Mott insulator Ti$_x$O$_{2x-1}$(N$_y$) and is integrated onto a 2D MoS$_2$ channel. The interplay of partial polarization switching in the ferroelectric LHO layer and reversible field-tunable metal-insulator transition (MIT) in Ti$_x$O$_{2x-1}$(N$_y$) layer concomitantly triggers polar to non-polar phase transition in the LHO layer between 200 and 220 K. This results in distinctive step-like features in the channel current during DC measurements, and random current fluctuations in high-speed measurements with slim anticlockwise hysteresis. Our devices show subthreshold slopes as steep as 25 mV/dec at 210 K, breaking the Boltzmann limit. Our gate stack is also potentially tunable for operation at temperatures of interest, presenting innovative gate stack engineering approaches for low-power computing solutions.

# Introduction

The demand for low-power computing solutions is increasingly critical in modern electronics, driving the need for innovative material and device designs that combine high-performance with minimal energy consumption. The increasing demand for energy efficiency is emphasized in the International Roadmap for Devices and Systems (IRDS), which identifies ferroelectric materials and their devices, as well as 2D channel materials, as promising solutions [1].

The inherent polarization states of ferroelectric materials can be programmed using an external electric field, offering tremendous opportunities in electronics[2]. Ultra-low power devices such as ferroelectric field-effect transistors (FE-FETs)[3,4] and negative capacitance FETs (NC-FETs)[5–7] harness these switchable polarization states to enable non-volatile memories and steep-subthreshold FETs. Progress on ferroelectric materials and their applications have paved the way for disruptive circuit topologies and novel computing architectures, spanning nonvolatile memories (NVMs), logic-in-memory (LiM) computing, the synthetic replication of biological elements, such as neurons and synapses, and in artificial neural networks (ANNs).[8,9]

Among the ferroelectric materials, doped hafnium oxide (DHO), with its unconventional ferroelectricity, has emerged as a key player due to its compatibility with complementary metal-oxide-semiconductor (CMOS) technology. DHO has revolutionized the landscape of ferroelectric materials by offering robust ferroelectricity at nanoscale thicknesses, non-toxicity, and the potential for large scale integration within CMOS technology.[10–12] DHO has been successfully integrated into device architectures for NC-FET devices offering very steep subthreshold slopes, enabling low operating voltages for low-power applications[6,13,14]. The dopants used can also determine the design parameters, such as capacitance matching, for these NC-FETs[15]. Additionally, DHO based FE-FETs have been reported with large memory window and long retention time[16–18]. Reports have also highlighted read-disturb effects in FE-FETs which manifest as steep-switching during slow readouts[19]. DHO also finds wide use in neuromorphic computing where partial polarization switching emulates synaptic behavior[20–22].

Precise gate-stack engineering is essential for integrating $HfO_2$ based ferroelectrics to develop both generalized and specific FETs for various applications.[23–25] With its many competing phases, the polymorphic nature of hafnia offers unique device behaviors when used in FETs[26,27]. A particularly interesting gate-stack is Mott insulator $Ti_xO_{2x-1}(N_y)$ as a bottom electrode (BE), in contact with DHO[28]. Proximity coupling between these layers through electrostatic and elastic boundary conditions results in reversible polar to non-polar phase transitions in DHO layer concomitant with metal-insulator transition (MIT) in the BE layer. Such induced phase transitions generate giant negative pyroelectric and electrocaloric effects as reported elsewhere[28]. The tunability of MIT through strain and composition also means that the phase transitions in DHO layer are tunable and can be engineered to temperatures of interest. It is also important to note that these phase transitions are both field and temperature dependent. Although both phase transition materials (phase-change FETs[29–31]) and ferroelectrics are independently explored in FETs, the potential for a coupled system in the context of low power FETs remains elusive.

On the other hand, 2D materials such as $MoS_2$ are the ideal channel materials for low-power FETs offering excellent electrostatic control even in ultra-scaled devices[1,32]. 2D FETs display high ON-OFF ratio, along with good mobility and ON currents.[33,34] Their Van der Waal layered nature allows easy integration in heterostructure devices including ferroelectrics[35]. They are actively pursued for applications in both conventional CMOS and emerging neuromorphic platforms[36–38]. 2-D materials like $MoS_2$ also provides the key benefit of increased surface electron density, vital to sustaining the ferroelectric polarization[39].

Here we propose and demonstrate a phase change ferroelectric FET (PCFE-FET) on few layer 2D $MoS_2$ channel, where the gate stack exploits proximity coupling effects between Mott insulator, $Ti_xO_{2x-1}(N_y)$ (phase-change layer) and ferroelectric layer (LHO). The metal-semiconductor-insulator phase transitions of the phase-change layer induce reversible polar to non-polar transitions in the LHO layer, which when coupled with partial polarization switching of LHO displays unusual step like behavior in quasi static $I_D$-$V_{BG}$, and random current fluctuations in high-speed measurements. The responses in the quasi-static characteristics show steep-switching activity, surpassing the Boltzmann limit near the transition temperatures, which are also tunable with strain and composition engineering. Using innovative gate stack design, our work demonstrates enhanced electrostatic control and pathways for future scaled technology nodes, and insights into designing low-power FETs for emerging computing technologies.[1]

# Results and Discussions

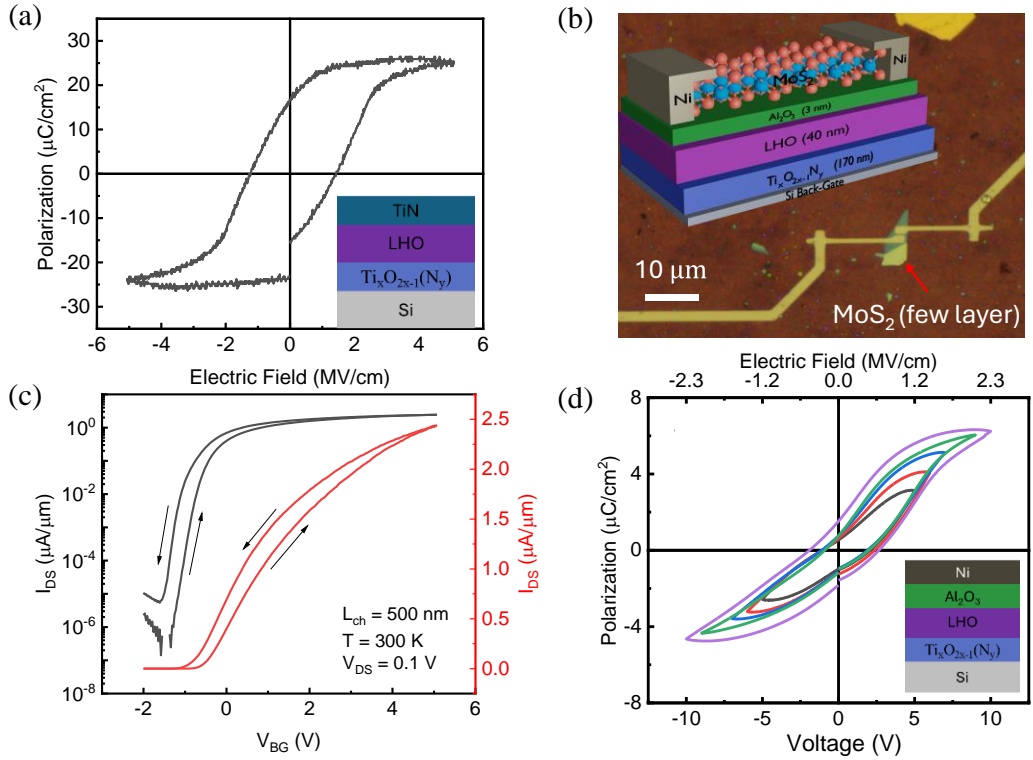

Fig. 1. Design of a PCFE-FET. a) Polarization vs. electric field on pristine LHO with TiN top electrodes, b) the optical image of the PCFE-FET, (inset) the schematic of the device, c) the quasi-static DC transfer characteristics of the PCFE-FET at room temperature, d) the polarization measurements on the device stack (shown in inset) showing sub-loops.

Lanthanum (7%) doped Hafnium oxide (LHO) was synthesized by chemical solution deposition and subsequently crystallized on a TiN deposited over p+ ($N_A = 10^{20}$ cm$^{-3}$) silicon via RF sputtering (methods). TiN layer oxidizes during the crystallization step into a Magneli phase of $Ti_xO_{2x-1}(N_y)$, which exhibits a tunable and reversible metal-to-insulator transition (MIT). These LHO/Magneli stacks display robust ferroelectric switching when contacted by TiN top electrodes at room temperature as shown in Fig. 1a, with a coercive voltage $V_C$ = 5.4 V ($E_C \cong 1.35\ MV\ cm^{-1}$), remnant polarization $P_r \cong 15.9\ \mu C\ cm^{-2}$, and saturation polarization $P_s \cong 24\ \mu C\ cm^{-2}$.

Next, we integrated the LHO/Magneli layer heterostructure into a 2D FET architecture, as shown in Fig. 1b (inset), where LHO layer is first capped with a thin $Al_2O_3$ interlayer (3 nm) via atomic layer deposition, onto which few-layer $MoS_2$ is transferred. Subsequently Ni source and drain electrodes are patterned, resulting in an n-channel FET[40]. Introducing $Al_2O_3$ dielectric interlayer is known to improve the poor interface between typical ferroelectrics and semiconducting channels resulting in better stability of the device[17,41]. Thickness of the interlayer was carefully optimized as 3 nm, which maximizes the device current and transconductance, while retaining signatures of ferroelectric polarization (Supplementary Fig. 1, Supplementary Note 1). The fabricated phase-change ferroelectric-FET or PCFE-FET (Fig. 1b) is a global back-gated transistor where the p+ doped silicon ($N_A = 1 \times 10^{20}$ $cm^{-3}$) contacting $Ti_xO_{2x-1}(N_y)$ acts as the back-gate electrode.

Fig. 1c depicts the variation of the drain currents ($I_{DS}$) of the PCFE-FET with the back-gate voltage ($V_{BG}$) at room temperature. The device shows anticlockwise hysteresis, which is typical of FE-FET devices due to the switching of polarization states.[35] To confirm the small anticlockwise hysteresis (~0.4 V) is indeed due to ferroelectric polarization, we measure P-E loops of the PCFE-FET device stack independently, by applying bias on $Ti_xO_{2x-1}(N_y)$ bottom electrode, and grounding the Ni source electrode. The ensuing FE capacitor (inset - schematic) and P-E loops are shown in Fig. 1d. Polarization saturation was observed when the device was tested with $V_{max}$ = 10 V, with coercive voltage $V_c \sim$ 2.5 V. However, when tested at lower $V_{max}$ (5, 6, 7, and 9 V) we observe minor loops arising due to partial polarization switching. The diminished values of $P_r$ and $P_s$ compared to Fig. 1a can be attributed to the differences in the top electrode, and the application of much lower electric fields.[42] The P-E loops combined with the transfer characteristics suggest that our PCFE-FET operates at sub coercive field, in the partial polarization switching regime, showing low $P_r$ and thus presents a low memory window and slim hysteresis (Supplementary Note 2)[3]. In the following, we do not consider the memory characteristics of PCFE-FET devices, but rather investigate its behavior in the partial switching regime.

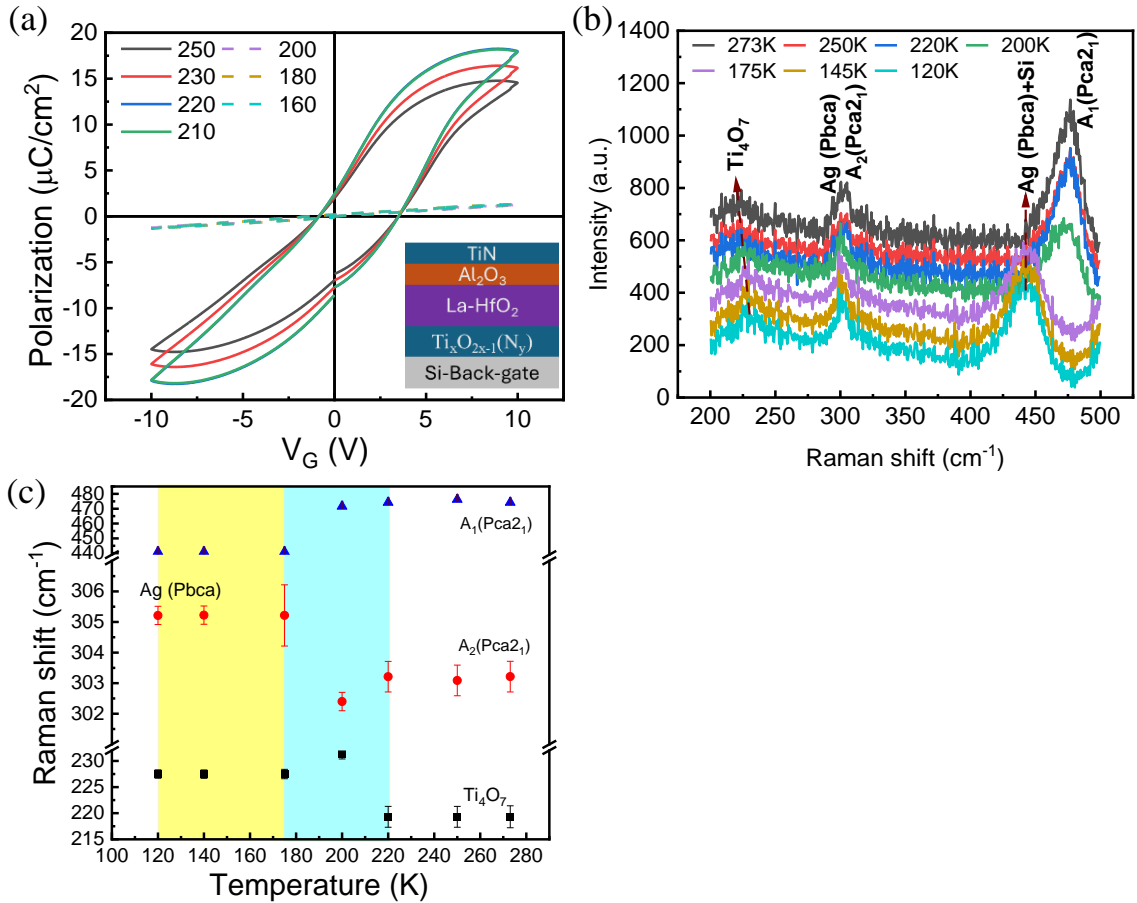

Fig. 2. Interplay of phase transitions and ferroelectric switching. a) Temperature dependent PE hysteresis measurements on the device stack with TiN top electrodes, b) Raman spectroscopy of the Magneli and LHO layers at various temperature, c) evolution of the Raman peaks with temperature.

We investigated the coupled phase-transitions in the Magneli and the LHO layers using temperature dependent PE loop measurements and Raman spectroscopy. With TiN as the top electrode, the PCFE-CAP devices (schematic in Fig. 2a, inset) exhibit PE loops with a slight increase in $P_r$ upon cooling from 300 K up to 210 K (Fig. 2a). The PE loops suddenly disappear at 200 K and below. Raman spectra from the heterostructure is shown in Fig 2b. We follow the evolution of positions of $A_1$ and $A_2$ peaks of LHO appearing at 473 cm$^{-1}$ and 303 cm$^{-1}$ at room temperature, and these peaks fingerprint the Pca2$_1$ phase of LHO. In addition, we also follow the B peak position corresponding to the Magneli phase at 221 cm$^{-1}$, which corresponds to its metallic phase. All the peak positions obtained from Lorentzian fits of peaks shown in Fig. 2b are consolidated as a function of temperature in Fig 2c. In the temperature range from 273 K to 220 K, A$_1$ mode of polar orthorhombic phase (Pca2$_1$) is uniquely observed at ~473 cm$^{-1}$ (A$_1$ peak) with weak temperature-dependent variation, indicating phase purity in LHO.[26] At 200K, the polar A$_1$ modes shift towards a lower frequency, and below 200 K, the A$_1$ mode disappears and the anti-polar Ag mode of LHO, which is a fingerprint of the Pbca phase appears at 441 cm$^{-1}$.[26] A$_2$ mode of LHO also follows a similar trend suggesting a clear polar to non-polar phase transition below 200 K, consistent with the disappearance of polarization in the PE loop measurements. The back gate contact is a mixture of metallic Magneli Ti$_x$O$_{2x-1}$ phases with traces of N (detailed characterization can be found in ref.[28]) and shows characteristic peaks around ~221 cm$^{-1}$ in the temperature range 220 K-273 K which shifts to a higher frequency of 231 cm$^{-1}$ at 200 K, finally settling at 227 cm$^{-1}$ below 200 K. This is consistent with a metallic to semiconductor phase transition between 200 and 210 K, and semiconducting to insulating phase between 200 to 175 K in the Magneli layer.[28,43,44] MIT with similar phase transitions was reported for Ti$_4$O$_7$ (another Magneli phase) in a Ti$_4$O$_7$/LHO stack, albeit with slightly lower transition temperatures: metal-to-semiconductor transition below 140 K and a semiconductor-to-insulator transition below 125 K.[28] Despite similar processing conditions as those in ref.[28], the increase of these transition temperatures in our samples points out to the effects of oxygen exchange and electrostatic boundary conditions imposed on the stack by the Al$_2$O$_3$ layer, suggesting that the transition temperatures are tunable. These results show that concomitant with the MIT transition of the Magneli layer, a polar to non-polar phase transition is proximity induced in the LHO layer. Such proximity coupling is governed by turning on (and off) the depolarizing field in the LHO layer below (above) the Metal-Semiconductor phase transition temperature, as well as the sudden changes in the elastic boundary conditions at the LHO- Magneli interface at these temperatures[28]. The effects of multiple phase transitions, field-dependent transition temperatures, and phase mixtures determine the behavior of the PCFE-FETs in the ~200 to 220 K temperature range, and we discuss this behavior next.

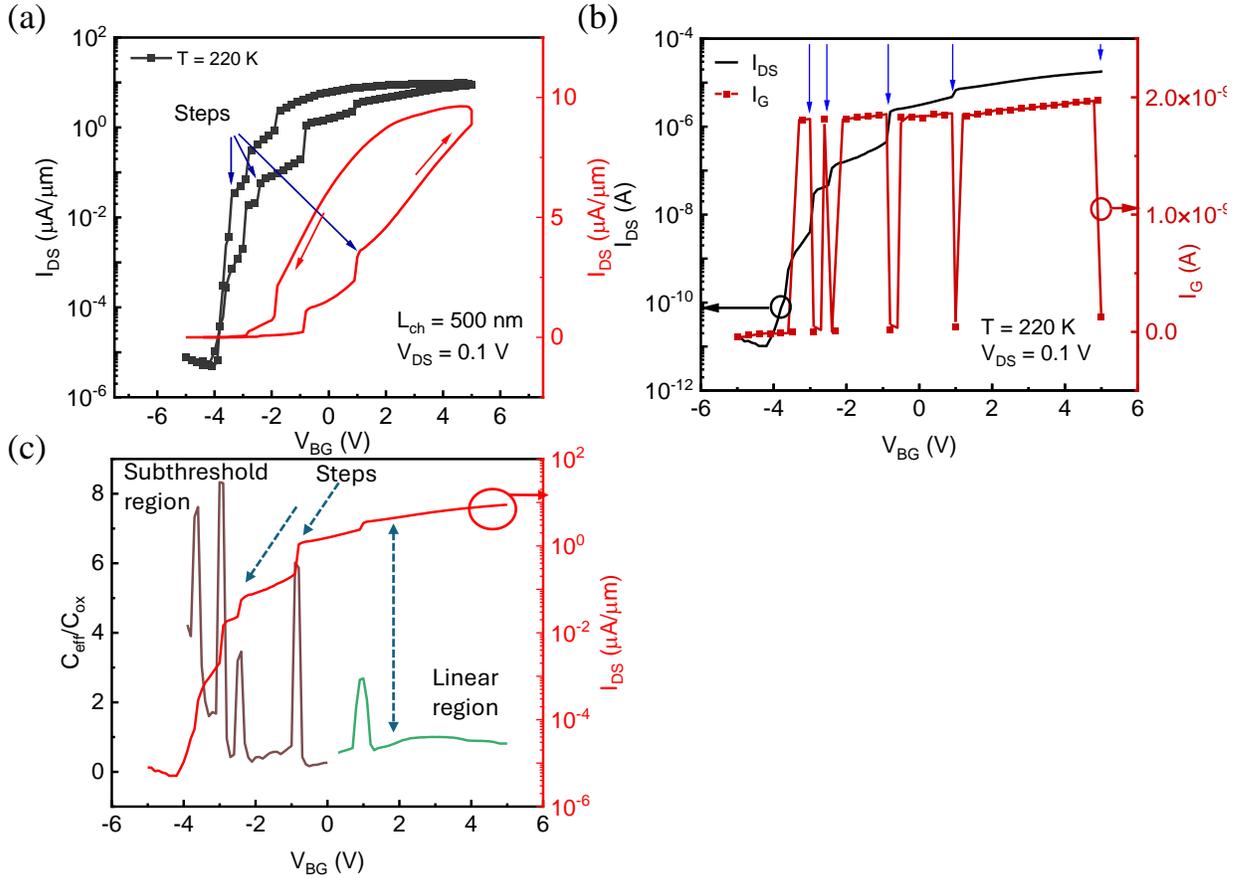

Fig. 3. Quasi-static begavior of PCFE-FETs. a) Peculiar step-like events in the transfer characteristics of the PCFE-FET in the range of 200 K to 220 K, b) correlation of the steps with fluctuations in gate leakage current, c) effective change in gate capacitance during the step-like events.

To understand the effect of phase transitions on the PCFE-FETs we conducted temperature dependent (120-300 K) quasi static DC measurements on our devices. In the temperature range between 200 and 220 K, on several tested devices (Supplementary Fig. 2) the drain current of the $MoS_2$ based PCFE-FET shows distinct and prominent step-like behavior, where the current shoots up over a single voltage sample and then tapers off, until the next event. These steps are visible both in forward and reverse sweeps, where the hysteresis direction of the transfer characteristic remains anticlockwise. Furthermore, these events are prominent in the subthreshold region of operations, although they do appear in the linear region as well. Similar steps do not appear below 175 K or above 220 K (Supplementary Fig. 3, 4).

Fig. 3b correlates the steps with the corresponding gate leakage during the measurement. Interestingly, the baseline leakage shows sudden decrease in gate currents ($I_G$) coinciding with these switching events. They revert to the baseline levels post the switching, indicating a transient event. The fluctuation in the $I_G$ strongly indicates the role of the gate stack in the step-like events in the transfer characteristics, and the temperature range of these effects strongly suggest the role of semiconductor/insulator to metal phase transition in the Magneli layer, and corresponding non-polar to polar transition in LHO layer. This is also supported by our experiments on $MoS_2$ FETs with e-beam evaporated amorphous $HfO_2$ dielectric as the gate oxide, where no peculiar features are visible at these temperatures (see Supplementary Fig. 5).

We estimated an effective gate capacitance change during these events to further glean the gate-stack characteristics responsible for step-like drain current behaviour. This is shown in Fig. 3c, where standard MOSFET equations have been used to extract an effective gate oxide capacitance $C_{eff}$ (see Supplementary Note 3). The $C_{eff}$ spikes during these events are consistent with the phase transitions in the gate oxide. Independent impedance measurements were also conducted on larger area electrodes (see Supplementary Fig. 6) where the magnitude and phase of the gate stack impedance show random fluctuations when the back-gate voltage is varied.

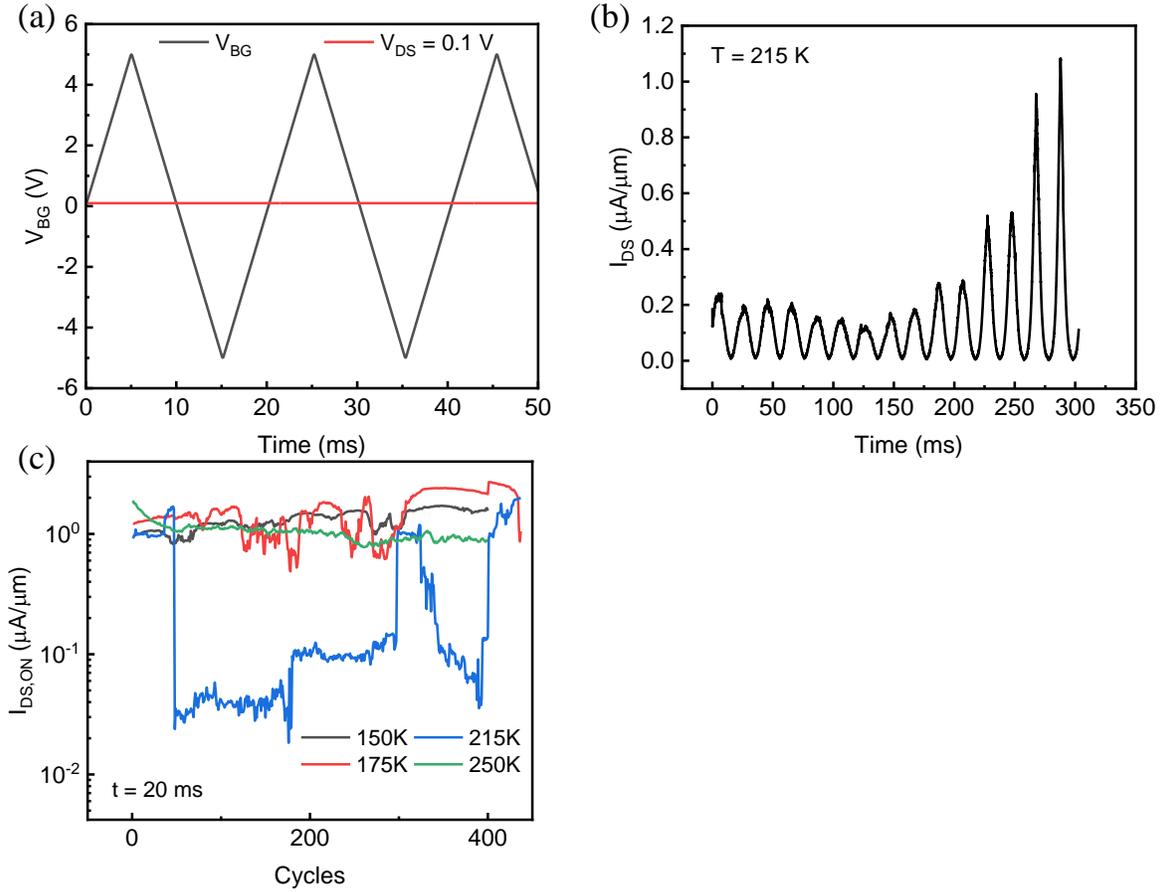

Fig. 4. High speed response of the PCFE-FETs. a) the input triangle gate voltage, b) the variation of drain current in the PCFE-FETs in response to the gate pulse in (a), c) the variation in ON current (at $V_{BG}$ = 5 V) of PCFE-FET at different temperatures, extracted from high-speed measurements in (b).

To further understand the origin of multi-step events in quasi static measurements, we performed pulsed (high speed) measurements on the devices as a function of temperature. The global-back-gated device geometry prevents us from applying very fast pulses due to the large gate parasitic capacitance. Hence, we choose a timescale of tens of milliseconds to probe our devices. The gate is stimulated with a triangle voltage (-5 to + 5 V, 20 ms period) at a constant drain bias as shown in Fig. 4a and the drain current is measured. At 215 K, in contrast to the step-like features visible in the DC transfer characteristics, high-speed measurements reveal random variations in the ON currents of the PCFE-FET (Fig. 4b). Furthermore, no discernible trends or periodicity is visible in these variations as evidenced by data shown in Supplementary Fig. 7, where multiple cyclic measurements at 215 K are shown. Interestingly, the high-speed measurements at 300 K remain stable with no variations present upon cycling the device (Supplementary Fig. 8), with consolidated $I_D$ vs $V_{BG}$ plots shown in Supplementary Fig. 9. We compare the maximum drain current in each sweep across various cycles and at a temperature range from 150 K to 300 K and present the data in Fig 4c (~400 cycles) and Supplementary Fig. 10 (~2000 cycles). Maximum fluctuations are visible around 215 K, while they start to die down below 175 K. Very stable characteristics are visible at temperatures less than 175 K and 250 K and above (Fig. 4c and Supplementary Fig. 10). It may be noted that the variations in the device currents are not correlated directly with the threshold voltage or the hysteresis. Simultaneous measurement of gate current in every voltage cycle shows an overall constant displacement (capacitive) current interspersed by kinks in some cycles, corresponding to gate stack capacitance changes, a consequence of phase transitions and polarization changes in the gate stack (Supplementary Fig. 11). High-speed measurements were able to temporally split the effect of these fluctuations into different cycles, also revealing that the time scale of these phase-transition events is slower than tens of milliseconds (Fig 4b, 4c).

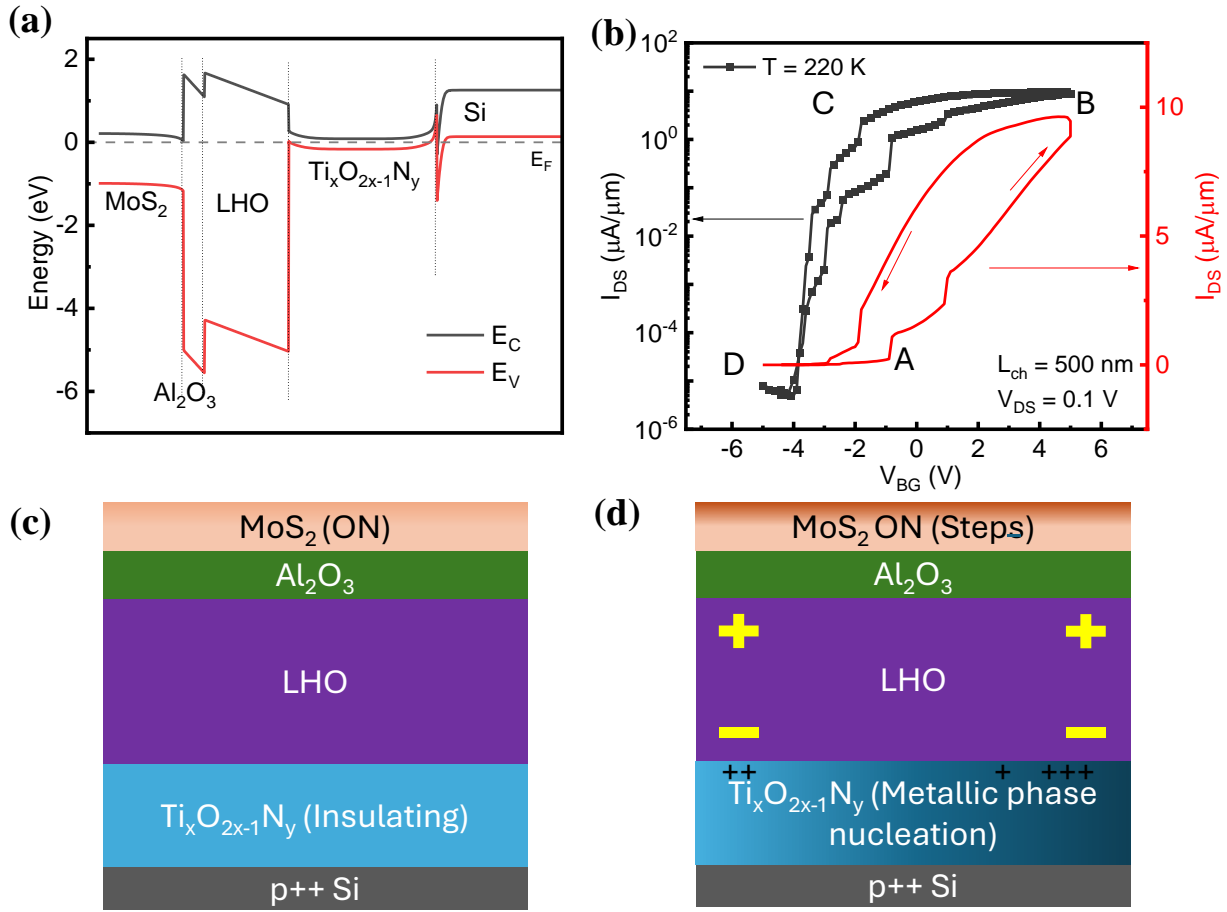

Fig. 5. Principle of operation. a) Equilibrium band diagram of the device's cross-section, b) the transfer characteristics of the PCFE-FET highlighting various regimes, c) the device cross section showing the insulating phase, d) the device cross-section showing nucleation of metallic phases.

Based on the structural and transport measurements described so far, we present the following comprehensive model to describe the device physics, especially in the 200 to 220 K temperature regime. Indeed, it is the combination of partial ferroelectric switching along with field tunability of the metal-semiconductor phase transition temperature in $Ti_xO_{2x-1}(N_y)$, and time dependent nucleation events of respective phases, which cause the peculiar steps in the quasi-static drain current in our PCFE-FET devices. The equilibrium band diagram of the device cross section along the gate stack is depicted in Fig. 5a, which is constructed using the material parameters provided in Supplementary Table 1. The $Ti_xO_{2x-1}(N_y)$ is in its insulating phase when the applied gate bias is 0, denoted by (A) in Fig. 5b (a simplified cross section is shown in Fig. 5c). For simplicity, the effect of polarization is not shown, which will cause additional band bending. Additionally, the bands are drawn assuming semiconducting nature in $Ti_xO_{2x-1}(N_y)$ [45]. Due to the large flat-band shift, MoS$_2$ channel is ON when $V_{BG}$ is 0 V. The Si back-gate and the $Ti_xO_{2x-1}(N_y)$ layer form a p+-n junction where the latter gets depleted of carriers.

When a positive $V_{BG}$ is applied, the depletion in the $Ti_xO_{2x-1}(N_y)$ layer reduces, resulting in a higher electron density in the $Ti_xO_{2x-1}(N_y)$ layer. It is known in Mott insulators that increase in electron density, i.e., electron doping, reduces the transition temperature [46–48]. Since we are operating on the verge of phase transition, reduction in $T_c$ nucleates metallic phase in the bottom electrode (Fig 5d). Additionally, electric field aids in the nucleation events of metallic clusters.[49–52] In thin films, phase transitions proceed typically through multiple nucleation events, which are distributed over time (analogous to Nucleation Limited Switching models in ferroelectrics)[53,54]. Through proximity effect, this triggers the nucleation events of polar phase in LHO, which are oriented upwards (in the direction of the field), that suddenly adds polarization charges (n-type) onto the channel (which is in ON/accumulation regime), causing a sharp step like increase in the MoS$_2$ channel current, as visible between points (A) and (B). The randomness in the high-speed measurements (10 ms pulses) arises due to the kinetics of these phase transitions and nucleation events, being slower than tens of milliseconds. The band bending at the $Ti_xO_{2x-1}(N_y)$ interfaces further reduces with the nucleation of the metallic phase, injecting more electrons and eventually reaching to a fully metallic state at point (B) indicated in Fig. 5b.

Next, when $V_{BG}$ is decreased and becomes negative, $Ti_xO_{2x-1}(N_y)$ is further depleted of carriers, which increases $T_c$, nucleating more insulating phase. This results in loss of polarization in LHO, and correspondingly causes a sharp reduction in carriers in the channel, depletes it further, resulting in pronounced step-like drops in the channel current. At (D), the Magneli layer is completely insulating, MoS2 is fully depleted, and the device is turned OFF. From (D) to (A), at large negative bias, the nucleation of metallic phase in the magneli layer can be triggered by the injection of electrons via tunnelling (due to the large band bending and low offsets to electron injection). The process repeats due to this back-and-forth cycling between metallic-insulating phases in $Ti_xO_{2x-1}(N_y)$, concomitantly triggering polar-non-polar transitions in LHO.

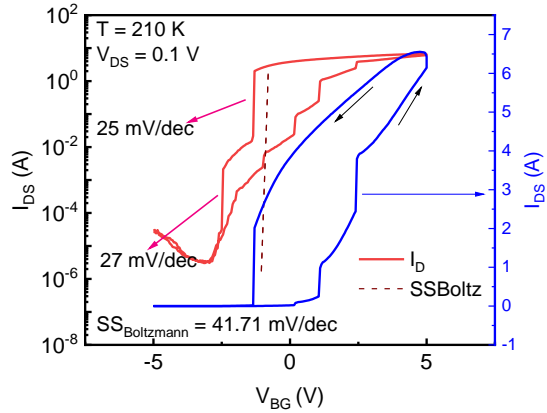

Fig. 5. Application of the PCFE-FET as a steep switching transistor surpassing the Boltzmann limit

Various applications of this switching phenomenon can be conceived by tailoring the properties of the gate stack and speed of operation. The step-like switching events can be utilized for a steep-switching FET, where the subthreshold slope can surpass the Boltzmann limit. Our PCFE-FET displays steep-switching activity when it is operated at the onset of phase transitions. Quasi-static DC measurements at 210 K yielded a minimum subthreshold slope of 25 mV/dec, much lower than the expected Boltzmann limit of 41.7 mV/dec at this temperature (Fig. 6), making this a potential approach towards low-power FETs.

We note that several phenomena can manifest similar to the step-like events in drain current visible in our devices. A recent report, by Mulaosmanovic et al. [19], shows an apparent steep-slope behavior in FDSOI based FEFETs where time-voltage dependence of ferroelectric switching can result in single or multiple steep-slope switching events in the FET device during the reverse sweep. This behavior was attributed to read-disturbs due to partial/full switching of polarization states, arising from the quasi-static nature of DC measurements, and displayed a disturb-free memory window in high-speed measurements.[19,55]

Although the steep switching in our PCFE-FETs has similarity to this apparent steep-switching, a slim hysteresis or lack of a clear memory window (Supplementary Fig 9), renders our gate-stack engineering approach as a controllable and desirable option for practical applications in low-power FETs. It may be noted that the transition temperatures in Mott insulators are tunable by stress, doping, stoichiometry and defects, and thus these effects are extendable to room temperature (or a desired temperature) by choosing and engineering Mott insulators with transitions close to these temperatures.

The kinetics of these phase transitions place a limit on the frequency of our PCFE-FETs, which is in the order of 10s of milliseconds in this work. Scaling the PCFE-FETs is a potential solution to counter the slow switching kinetics, and our platform with hafnia-based gate and 2D material channel is indeed scalable[56]. However, the slow(er) kinetics that result in fluctuations in the high-speed measurements can be exploited for systems such as random number generators where the stochasticity of the nucleation of metallic/insulating phase can be leveraged[57]. Additionally, the phase transitions in these devices can be utilized for added non-linearity in applications such as reservoir computing where FEFETs are popular[9].

To conclude, we demonstrate a PC-FEFET which is a uniquely designed gate stack architecture integrated onto a 2D channel. The gate architecture offers concurrent phase transitions along with partial ferroelectric switching, resulting in devices that show steep subthreshold slopes beating the Boltzmann tyranny. Near the coupled phase-transition temperatures, which are tunable, these devices show unique features including random step-like events, high-speed fluctuations, and steep-subthreshold switching. Our gate stack engineering, combined with energy efficient ferroelectric hafnia and 2D material platforms, holds a great potential for low-power FETs, and emerging paradigms such as random-number generators and reservoir computing.

## Methods

**Material Synthesis and Device Fabrication**: La-doped hafnia (LHO) films were prepared by a cost-effective chemical solution deposition reported in detail elsewhere[28]. Hf(IV)-2,4-pentanedionate (Alfa Aesar, 97% purity) and Lanthanum (III) acetylacetonate hydrate (Alfa Aesar, 97% purity) were used as the ingredients for preparing the precursor sol. The cationic La % in the precursor was selected as 7%. We use TiN (170 nm, RF sputtered)/Si as the bottom contact, and the solution containing La & Hf cations was spin-coated on top of the TiN/Si stack and pyrolyzed at 350 °C. The process was repeated to obtain the desired thickness (40 nm LHO at n=12). The final films were annealed at 680 °C to crystallize the LHO. The annealing process also oxidizes the TiN to its Magneli form $Ti_xO_{2x-1}(N_y)$.

For further FET device fabrication, a thin layer of $Al_2O_3$ (3 nm) was deposited by atomic layer deposition (ALD Beneq) using trimethyl aluminium (TMA) and water as precursors at 250 °C. $MoS_2$ flakes (Graphene Supermarket) were mechanically exfoliated onto this sample using polydimethylsiloxane-based stamps (Gel-Pak) and scotch tape. Acetone, isopropyl alcohol (IPA), and deionized water (DI) were used to clean the sample between each process. Optimum flakes of thickness 5 – 7 nm (few-layer $MoS_2$) were identified optically. Two-step Electron-beam lithography (Raith Eline/Pioneer) was used to define source and drain contacts, with a channel length of 500 nm. A contact metal of Ni (50 nm) was deposited by electron-beam evaporation (Leybold) and lifted off, completing the device fabrication.

**Device Characterization:** Electrical measurements were carried out using the Keithley 4200 SCS parameter analyzer in vacuum using a Lakeshore cryogenic probe station in the range of 77 K to 300 K. Pulsed, high-speed measurements were carried out using the Keithley 4225 ultra-fast pulse measurement unit (PMU) included in the parametric analyzer.

Raman spectra were obtained using the LabRAM HR (UV) system with a 532 nm laser excitation source from Horiba. Temperature control during Raman measurement was achieved using the HFS600E-PB4 Probe Stage from Linkam Scientific Instruments.

## Data Availability

The data that support the findings of this study are available from the corresponding author upon reasonable request.

**Acknowledgment**

Major part of this work was carried out at Micro and Nano Characterization Facility (MNCF), and National Nanofabrication Center (NNFC) located at CeNSE, IISc Bengaluru. P.N. acknowledges Start-up grant from IISc, Infosys Young Researcher award, and ANRF-SERB and DST with grant numbers CRG/2022/003506 and INT/ROK/JRP/2024/394. The authors acknowledge Vishnu Kumar for his assistance in low-temperature electrical measurements.

**Competing Interests**

The authors declare no competing interests.

## Author Contributions

‡S.S and J.M.A contributed equally. S.S, J.M.A and P.N conceived the project. J.M.A prepared the LHO samples, conducted Raman and structural characterization of the devices; S.S made the devices, and conducted the transport measurements. S.S., J.M.A., P.N., and N.K.B discussed the data and its interpretations. S.S., P.N., co-wrote the manuscript with input from J.M.A and N.K.B.

All authors approve the final version of the manuscript.

# Additional Information

**Supplementary Information**

# Hafnia-based Phase-Change Ferroelectric Steep-Switching FETs on a 2-D MoS$_2$ platform

*Sooraj Sanjay[‡], Jalaja M. A[‡], Navakanta Bhat\*, and Pavan Nukala\**

Centre for Nano Science and Engineering, Indian Institute of Science, Bengaluru, India.
Email: navakant@iisc.ac.in, pnukala@iisc.ac.in

[‡]Centre for Nano Science and Engineering, Indian Institute of Science, Bengaluru, India.

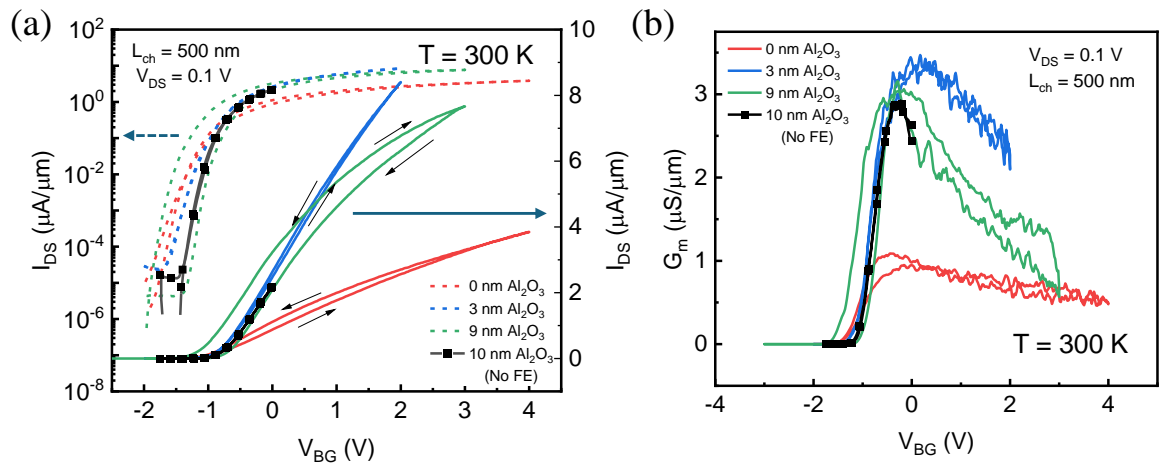

Figure 6. The effect of Al$_2$O$_3$ interlayer thickness on device performance, a) transfer characteristics, b) transconductance.

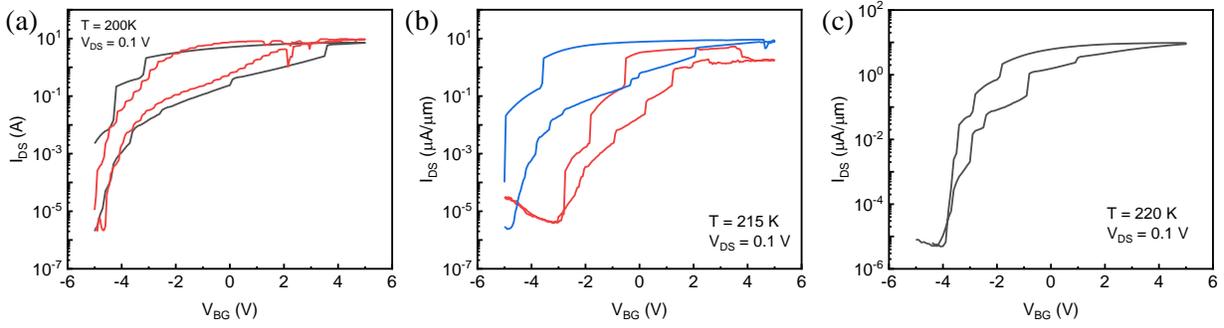

Figure 7. Transfer characteristics show step-like events. a) at T = 200 K, b) at T = 215 K, c) at T = 220 K. (Colors indicate different devices or measurements)

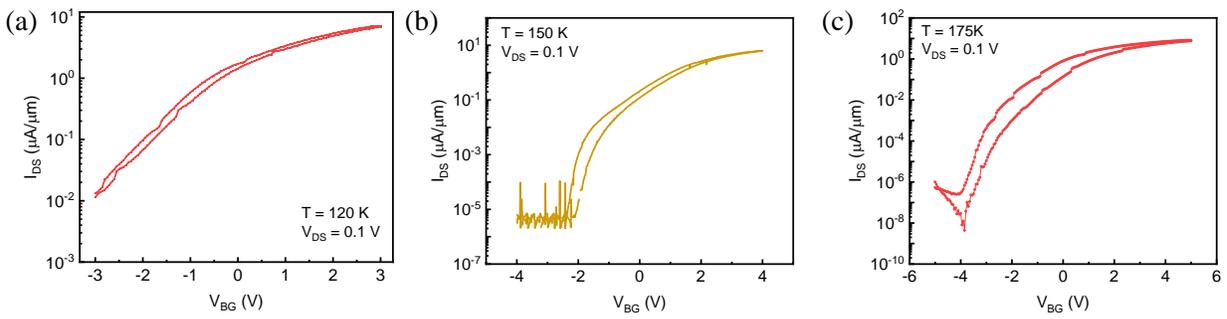

Figure 8. The response of the PCFE-FETs for T < 175 K, showing negligible step-like activity.

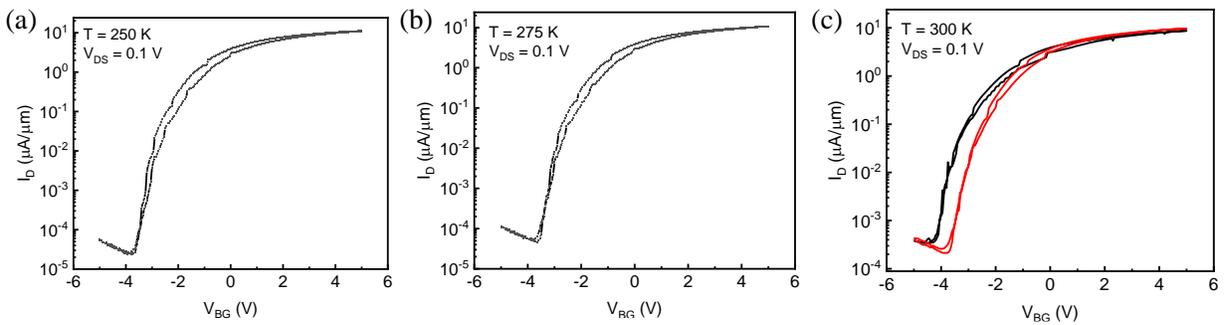

Figure 9. The response of the PCFE-FETs for T > 250 K, showing minimal step-like activity.

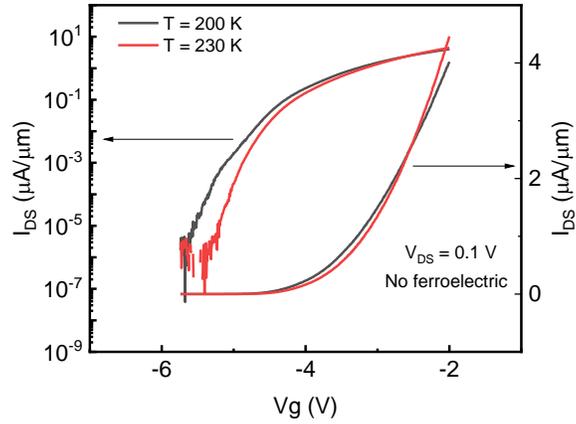

Figure 10. The transfer characteristics of a MoS$_2$ back-gated FET using non-ferroelectric (amorphous) HfO$_2$, showing no step-like activity.

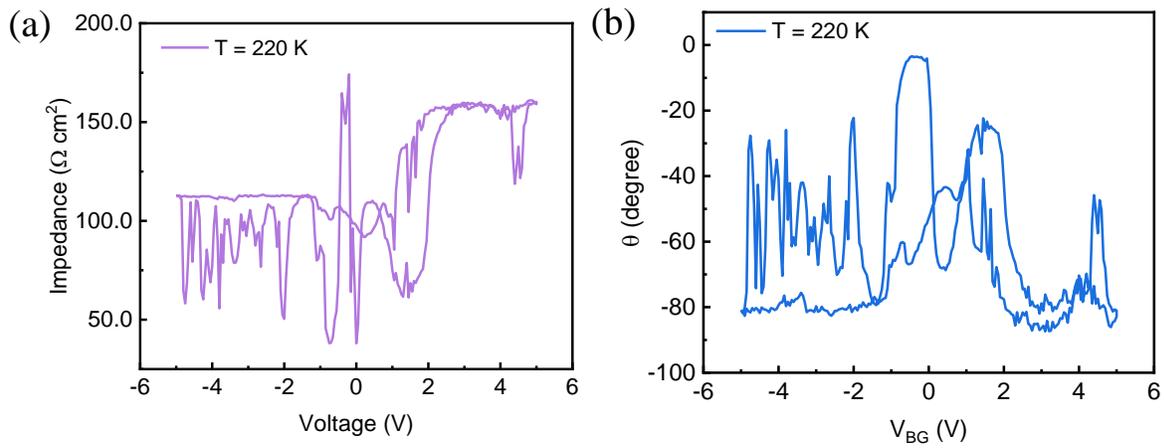

Figure 11. Impedance measurements on the device stack showing random fluctuations. a) Impedance vs. voltage, b) loss angle vs. voltage.

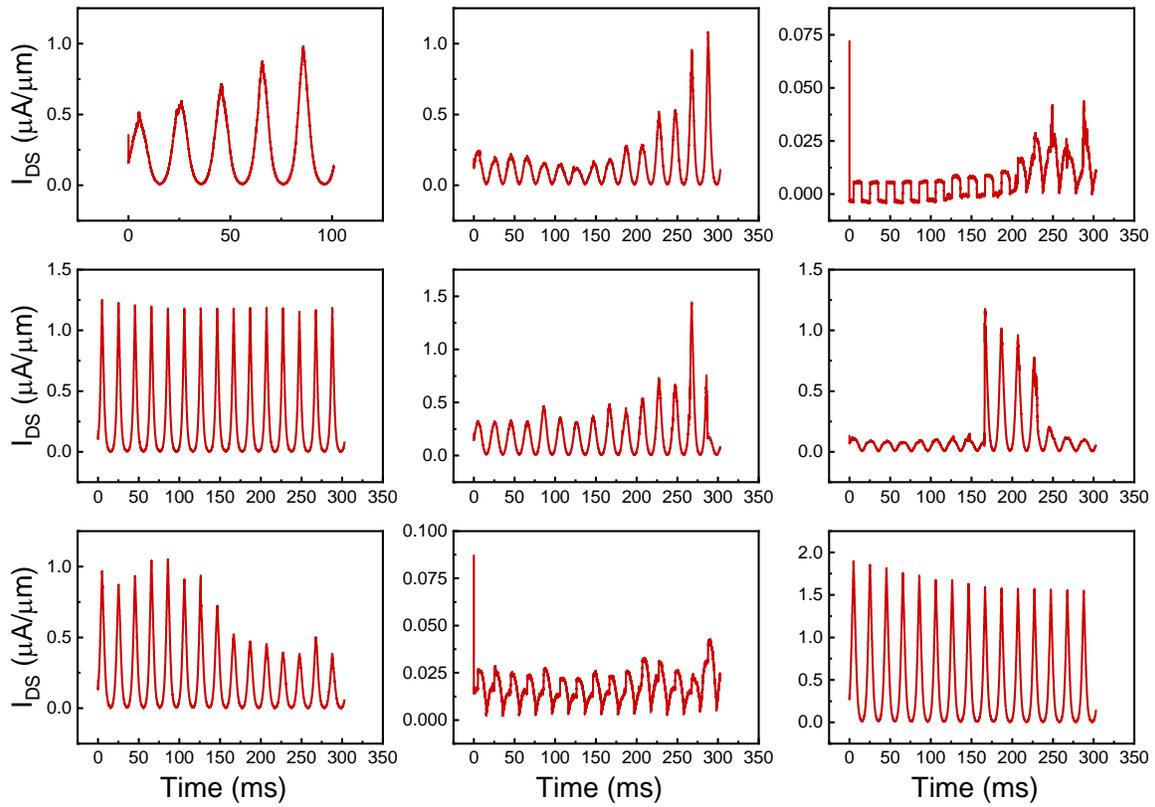

Figure 12. Various responses of the PCFE-FET with triangle gate pulse at T = 215 K, clearly showing random fluctuations in the drain current.

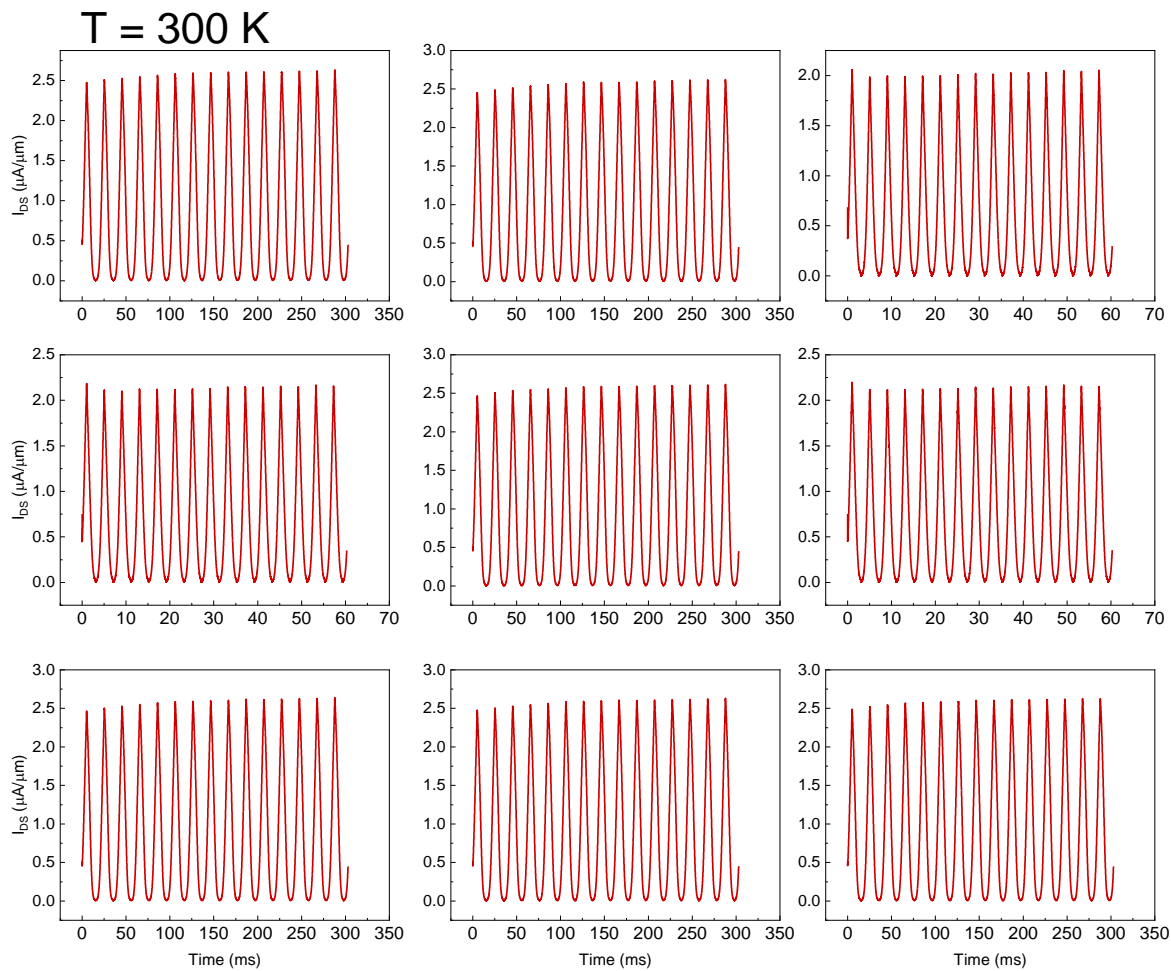

Figure 13. Stable drain currents visible for triangle gate pulses at T = 300 K

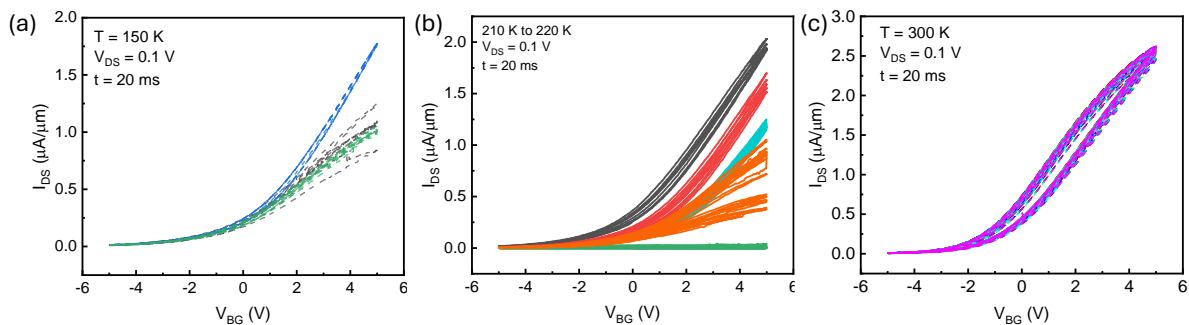

Figure 14. The transfer characteristics of the PCFE-FET extracted from high-speed triangle gate bias measurements. A) small variations in drain current at T = 150 K, b) very large deviation from mean behavior visible between 210 K and 220 K, c) very stable device behavior at T = 300 K.

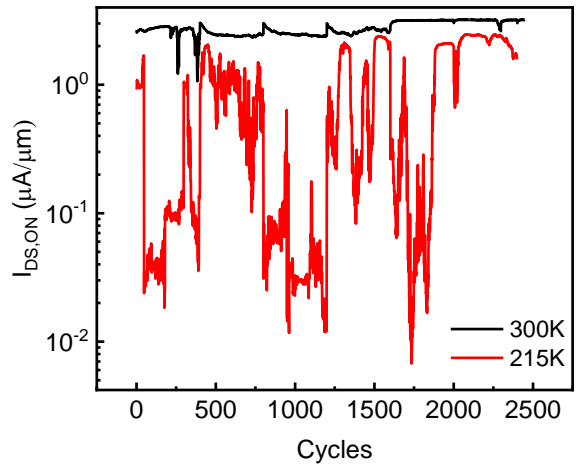

Figure 15. The comparison of cycle-to-cycle variation in ON current ($V_{BG}$ = 5 V) between T = 215 K and 300 K

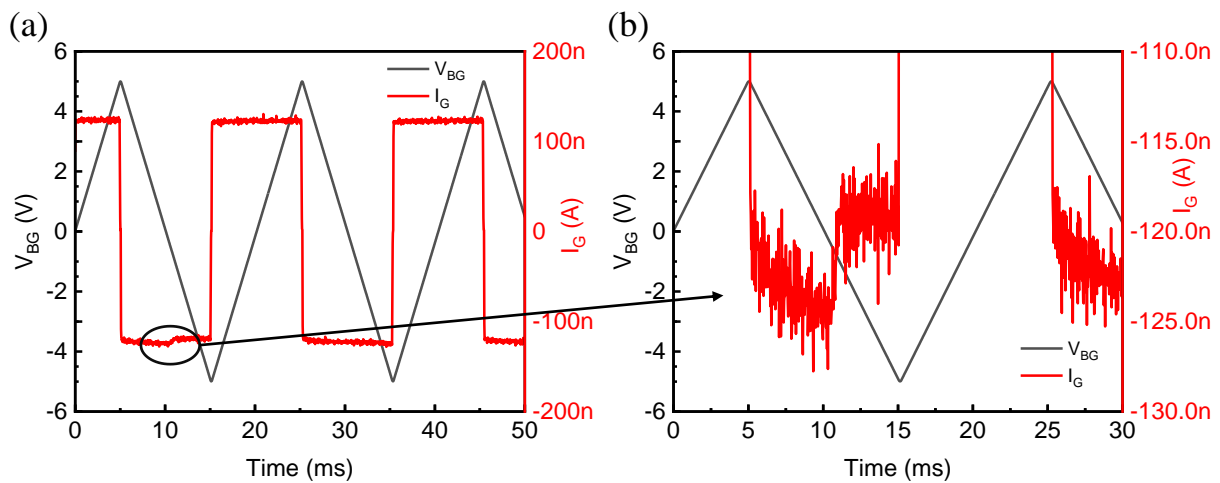

Figure 16. Gate current in high-speed measurements a) small kinks in gate current which are indicative of phase transitions and polarization changes, b) enlarged section of (a) highlighting a switching event.

**Supplementary Notes**

Supplementary Note 1:

The interface between the semiconductor and the insulator is a site for many defects, collectively called interface traps. Interface traps capture charges from the channel, reducing the mobility and drain currents and resulting in a clockwise hysteresis for the transfer curve. On the contrary, an anticlockwise hysteresis is obtained when there is ferroelectric switching due to the polarization charges which shifts the threshold voltage and raises the channel current. In an FE-FET, there is a competition between the polarization charges from the ferroelectric and the interface traps. Reports have suggested poor interfaces and large interface traps when a ferroelectric is directly in contact with the semiconductor.[1,2] This is evident in Supplementary Figure 1, where lower currents and transconductance are obtained without an interlayer. The current levels are improved with the introduction of a thin $Al_2O_3$ layer. However, the presence of the interlayer diminishes the ferroelectric switching and reduces the anticlockwise hysteresis window. Although the device current and peak transconductance improves with $Al_2O_3$ thickness, the anticlockwise hysteresis narrows and turns clockwise at higher interlayer thicknesses. The competition between polarization charges and interface traps indicates an optimum thickness of $Al_2O_3$ where the ON currents and transconductance are maximized, while still displaying signatures of ferroelectric switching (i.e., anticlockwise hysteresis). At 3 nm of $Al_2O_3$, the hysteresis is anticlockwise, and the device currents are identical to using only $Al_2O_3$, which is considered optimal for this application. Beyond 3 nm of $Al_2O_3$, the device shows clockwise hysteresis due to dominant trapping effects and diminished polarization.

Supplementary Note 2:

The typical memory window (MW) of an FE-FET can be given by the simplified expression ($MW = 2\alpha E_C t_{FE}$), where $0 < \alpha < 1$, and $t_{FE}$ is the ferroelectric thickness.[3] Although this calculation predicts a window as high as 10.8 V for our devices, the bias needed to switch this device would be even higher due to the voltage drop across the $Al_2O_3$ interlayer. We operate the PCFE-FET in the sub-coercive field, where the memory window is smaller.

Supplementary Note 3:

In the linear region, the standard equation (1) given below is used to calculate $C_{eff}$, where $\mu_{FE} = 30 \text{ cm}^2/\text{Vs}$ is obtained from FETs with $Al_2O_3$ as dielectric, W, L are device dimensions, $V_{DS}$ is the drain to source bias, and $V_{BG}$ is the back-gate bias.

$$g_m = \frac{dI_{DS}}{dV_{BG}} = \mu_{FE} C_{eff} \frac{W}{L} V_{DS} \tag{1}$$

In the subthreshold region, we use the below equation (2) over the envelop of the transfer characteristics, where k is the Boltzmann constant, and T is the temperature; $I_{DS0}$ is the current at $V_{BG} = V_{th}$, n is the subthreshold slope parameter is related to $C_{eff}$ as shown in (3), $C_{MoS2}$ is the depletion capacitance from the $MoS_2$ channel.

$$I_{DS} = I_{DS0}\, e^{\frac{q(V_{BG} - V_{th})}{nkT}} \tag{2}$$

$$n = 1 + \frac{C_{MoS_2}}{C_{eff}} \tag{3}$$

Supplementary Table 1:

| Material | Electron Affinity ($\chi$, eV) | Bandgap ($E_g$, eV) | Carrier Density (approx. $N_D$, $N_A$, cm$^{-3}$) | Ref |
|---|---|---|---|---|
| Si | 4.05 | 1.12 | 1 x 10$^{20}$ | 4 |
| Ti$_x$O$_{2x-1}$(N$_y$) (Ti$_4$O$_7$) | 2.64 | 0.25 (semiconducting) | 7.81 x 10$^{18}$ (semiconducting) 3.125 x 10$^{20}$ (metallic) | 5–8 |
| LHO | 2 | 5.95 | - | 9,10 |
| MoS$_2$ | 4.2 | 1.2 | 1e16 | 11–13 |
| Al$_2$O$_3$ | 2.58 | 6.65 | - | 14,15 |